%Paper: nucl-th/9301005
%From: BALDO@CATANIA.INFN.IT
%Date: Thu,  7 JAN 93 15:42 GMT

%%%%%%%%%%%%%%%%%%%%%%%%%%%%%%%%%%%%%%%%%%%%%%%%%%%%%%%%%%%%%%
%
%  Plain TeX  version 3.0
%
%%%%%%%%%%%%%%%%%%%%%%%%%%%%%%%%%%%%%%%%%%%%%%%%%%%%%%%%%%%%%%
%
% Figures included at the end of the manuscript in a
% postcript file to be extracted from the text
%
%
%%%%%%%%%%%%%%%%%%%%%%%%%%%%%%%%%%%%%%%%%%%%%%%%%%%%%%%%%%%%%%
\magnification=\magstep1
\nopagenumbers
\footline={\ifnum\pageno>0
           \hss\tenrm\folio\hss
           \else\hfil\fi}
\overfullrule=0pt
\pageno=0
\baselineskip=20 pt
{\ }
\topinsert
\noindent
January 1993 \hfill  Catania University preprint n. 93/01
\endinsert
\vskip 1 true cm
\hrule
\vskip 4 true cm
\bigskip
\bigskip
\bigskip
\centerline{\bf Revealing intermittency in nuclear
 multifragmentation}
\centerline{\bf with 4$\pi$ detectors}
\bigskip
\centerline{M.Baldo, A.Causa and A.Rapisarda}
\medskip
\centerline{Universit\` a di Catania and INFN, Sez. Catania,}
\centerline{ Corso Italia 57, 95129-Catania, Italia}
\bigskip
\bigskip
\vskip 2 truecm
\hrule
\vskip 1 truecm
\noindent
{\it Submitted to Phys. Lett. B}
\vfill\eject
\bigskip
\bigskip
\par\noindent
{\bf Abstract}
\par
{\it The distortion on the intermittency signal, due to
detection efficiency and to the presence of pre--equilibrium
emitted particles, is studied in a schematic model of nuclear
multifragmentation. The source
of the intermittency signal is modeled with a percolating system.
The efficiency is schematized by a simple function of the fragment
size, and the presence of pre--equilibrium particles is simulated
by an additional non--critical fragment source.
No selection on the events is considered, and therefore all events
are used to calculate the moments.
It is found that, despite the absence of event selection,
the intermittency signal is quite resistant to the distortion
due to the apparatus efficiency, while the inclusion of
pre--equilibrium particles in the moment calculation
can substantially reduce the strength of the signal.
Pre--equilibrium particles should be
therefore carefully separated from the rest of the detected
fragments, before the intermittency analysis on experimental
charge or mass distributions is carried out.
}
\bigskip
\noindent
\vfill\eject
\bigskip
%\noindent
%{\bf 1. Introduction}
%\smallskip
\par
In high energy proton--nucleus and nucleus--nucleus collisions,
as well as in heavy--ion reactions at intermediate energies,
events with very
large charge particle multiplicities are observed. The theoretical
analysis of these events is hindered by the complexity of the
reaction mechanisms, and often biased by model dependent
assumptions. In the experimental data large fluctuations in
various physical quantities are apparent, either in each event,
either from one event to another.
In recent years it was realized [1-3] that these fluctuations
are quite useful for the study of the nuclear dynamics and the
reaction mechanisms. They can be analyzed
in terms of global variables, whose trends can be used to
characterize high multiplicity events in an essentially model
independent way. In particular, the intermittency analysis [2,3]
has proven to be one of the most powerful and promising method to
this respect. Intermittency patterns have been experimentally
observed, both for rapidity charge particle distributions in high
energy proton--nucleus collisions [2], and for fragment charge
distributions in nuclear emulsion data on projectile
multifragmentation in peripheral heavy--ion collisions at
energies around 1 GeV/A [4].
\par
The intermittency analysis of the fluctuations
is essentially a multifractal analysis of a given distribution.
This method has been developed in many fields of physics, ranging
from hydrodynamics to astrophysics.
For the problem in exam, the intermittency analysis is performed
by studying the behaviour of the moments of the distribution, as
one varies the resolution $\delta s$ with which the distribution
itself is considered. Only event to event fluctuations
will be considered in this work, therefore the definition of the
so called scaled factorial moments $F_i$ which will be used, reads
$$
 F_i \, =\, {\sum_k \langle n_k(n_k -1)\ldots
 \ldots (n_k -i+1)\rangle
                 \over {\sum_k \langle n_k \rangle^i}}
\eqno(1)
$$
\noindent
For a fixed resolution $\delta s$, the interval of $s$ values is
divided in bins of size $\delta s$. Then
 $n_k$ is the number of fragments in the $k$--th bin for a given
event, and the average is performed on the considered set of
events. A diverging power law behaviour of these moments with
decreasing resolution is, by definition, the intermittency signal,
namely $F_i \sim {(\delta s})^{-\lambda_i}$, $\delta s\,
\rightarrow 0$. In a  plot of $\log F_i$ versus $-\log \delta s$,
this corresponds to the linear rise of the logarithms of the
factorial moments. The corresponding slopes $\lambda_i$ are called
the intermittency indices, one for each order $i$.
For a discrete and limited distribution, like the mass or charge
distributions, this behaviour is of course demanded only in the
limited range of physically admissible values of $\delta s$,
in particular $\delta s$ cannot be smaller than 1.
\par
In this work we consider the problem of the
disturbances on the intermittency signal which
can be produced either by the inefficiencies of the detection
apparatus, either by the reaction mechanism. In fact, the
detection of the intermittency signal could be strongly reduced,
or even suppressed, by the finite efficiency of the
detection apparatus. This is of particular interest for the
experimental works with 4$\pi$--detectors for heavy--ion collisions
at intermediate energy, which have been recently put in
operation or are planned for the near
future. A similar analysis has been presented in ref. [5] for other
global dynamical variable, like the transverse momentum,
the flow angle, and so on.\par
On the other hand, the source of the intermittency signal is
believed to be associated with the
critical behaviour of the nuclear system [3], either due to a phase
transition or to a mechanical instability [3,6]. However, in
general one can expect that only a part of the nuclear system
reaches the critical region, and therefore that the intermittency
signal can be clearly detected only if
the fragments coming from the participating zone are properly
selected.
\par
The simplest model which produces an intermittency
signal, associated with a phase transition, is the percolation
model [3,7]. It has been widely used [8] in the analysis of the
experimental data on mass and charge distributions
in heavy ion reactions. In this work we assume that the source
of the intermittency signal in the mass/charge distribution
can be modeled by a percolating system, and we
study the effects on this signal of various detection
inefficiencies, as well as of the presence of another non--critical
source. The results of this analysis should not be dependent on the
particular model, which is used to generate the intermittency
signal.
\par
The apparatus inefficiency is schematized in a simple way, by
assigning to each fragment of size $s$ a probability $Prob(s)$
to be actually detected, and therefore a probability $1 - Prob(s)$
to escape undetected from the apparatus. In a given event generated
by the computer, for each cluster of the percolation model, which
is assumed to represent a nuclear fragment, a random number $x$,
belonging to the interval (0,1), is generated and compared
with $Prob(s)$. If $x \leq
Prob(s)$, the fragment is assumed to have been detected,
otherwise it is assumed to have passed undetected. This global
parameterization of the efficiency can, in particular, roughly
represent the geometrical
efficiency of the apparatus. Of course the particular form of the
probability distribution $Prob(s)$ depends, in this case, not only
on the geometry of the detector,
but also on the reaction dynamics, since in general fragments of
different size are emitted preferentially in different
directions, where the geometrical
efficiency can be rather different. However, in this paper we
are not interested in any particular detector or heavy--ion
reactions, but rather to establish general trends, which can be
used as a guidance to the
inefficiency effects that one can expect. Therefore, the function
$Prob(s)$ will be used to represent globally the overall efficiency
of the apparatus, including the geometrical efficiency,
the detection efficiency, the threshold cuts, and so on. \par
In the nuclear emulsion data [4], where the intermittency signal
was observed, the charge of each fragment of the projectile
multifragmentation was identified, and only the events for which
the total sum of the fragment charges was
equal to the projectile charge were included in the analysis.
Because of the nuclear emulsion efficiency, this selection can bias
the final result. However, simulations with a model [6], which
successfully reproduces the intermittency signal, show that
in these cases the bias does not affect the main trends. With 4$\pi$
detectors for heavy ion collisions at intermediate energy,
the average total efficiency
$\overline{Prob}$ can hardly exceed 0.7--0.8. With average charged
fragment multiplicities of the order 40 to 60, to maintain such
a strong selection of the events would imply a severe reduction
of the counting rate, even down to the limits of the experimental
feasibility, in the worst cases.
It is customary, for this reason, to apply a less stringent
selection on the events, usually by only demanding [5]
that the total sum of the observed fragment charges
is not less then 90\% of the initial total charge, or so.
In this paper we will adopt the extreme attitude of excluding any
selection on the events, so that all the events will
be used in calculating the moments. This means that, even if a
few fragments are missing in a given event after applying the
efficiency filter $Prob(s)$, the detected fragment distribution
will still be used for calculating the numerator and the
denominator of eq. (1) and the corresponding contribution of the
event to the average value. It is expected that the introduction
of a proper selection on the events would
result in an enhancement of the intermittency signal.\par
In the simulation we will consider, for simplicity, only
one type of site in the percolating system, and therefore the size
variable $s$ can be interpreted as the charge or the
mass of the fragment, according to the
problem under study. This oversimplification
does not affect the analysis that will be presented. \par
As a side remark, it has to be noticed that in computer
simulations much care has to be
taken with the random generator routine, which can present
undesired spurious correlations between calls.
The routine used in the present work was checked under various
statistical tests [9]. Furthermore, care must be taken about the
convergence of the results with the numbers of events.
The results here presented were obtained with the standard
number of events equal to 30000. We checked that for a larger
number of events the results are essentially unaffected, while
for a smaller number the results can show fluctuations.
\par
The intermittency signal can be generated in a
percolating system by choosing the values of the site probability
$p$ and of the bond probability $q$ just equal to the critical
values $p_c$ and $q_c$ for the corresponding infinite system,
in agreement with the general rule that intermittency
is associated with critical points. In general it is
considered more realistic to assume the parameters $p$ and
$q$ to vary randomly in some range of values, since
in the experimental situation events with different nature
are usually mixed together. In this case intermittency
appears whenever the range of allowed random values includes
the critical ones. This point is illustrated
in the case of a $6^3$ cubic lattice in Fig. 1,
where, for $p = 1$, three intervals of uniformly distributed
values
of $q$ are considered. As expected, the linear rise of the factorial
moments is present only for the interval $0.2 < q < 0.3$,
in agreement with the critical value $q_c = 0.23$, at $p =1$,
in the infinite system. This particular system will be
considered in the sequel as the source of the intermittency signal.
It is characterized by the intermittency indices
$\lambda_2 = 1.73\cdot 10^{-3}$, $\lambda_3 = 4.78\cdot 10^{-3}$
and $\lambda_4 = 1.02\cdot 10^{-2}$. The relevance of event mixing
was also emphasized recently in ref. [10], where intermittency
was observed only in a particular dynamical regime
of the expanding emitting source model [11] of
multifragmentation. In general mixing of events enhances
the intermittency signal, if it is present, by increasing
the values of the critical indices.
\par
The efficiency function $Prob(s)$ can be quite different for
different detectors. For some experimental apparatus
the efficiency is increasing with the size $s$, and it can
be very low at small $s$ values. The opposite situation is also
possible, with some limit value of $s$, above which the fragments
are undetected or unresolved. For a constant efficiency $Prob(s)
= \overline {Prob}$, the intermittency signal appears essentially
undisturbed, with a pattern hardly distinguishable from the one
of Fig. 1b. We have checked that, even with a $\overline {Prob}$
value as small as 0.5, the linear rise, and the corresponding
slopes, remain mainly unaffected, despite the fact that
on the average only 50\% of the fragments are detected.
A slightly more realistic case is considered in Figs. 2a--c,
where the efficiency function is assumed to be of the form
$$
  Prob(s)\, =\, \cases {P_1 + {{1 - P_1}\over {s_0 - 1}}\cdot (s-1)
&if  $s\leq s_0$\cr 1 &if $s > s_0$\cr}
\eqno (1)
$$
\noindent
where $P_1$ is the efficiency for $s = 1$, and $s_0$ is the
value of $s$ above which one assumes $Prob(s) = 1$. The latter
has been fixed at $s_0 = 10$ in the calculations of Fig. 2. To the
extent that the value of $P_1$ is not too small, we have found that
the intermittency signal is only slightly disturbed, in a
wide range of $s_0$ values. Only when $P_1$ is close to zero,
the linear rise of the factorial moments is destroyed. These
results indicate that the detector efficiency must satisfy
only mild conditions, in order to be sensitive to the intermittency
signal. Essentially, in the specific schematic example, the condition
$P_1 \equiv Prob(1) \geq 0.2$ is required, without any stringent
restriction on $s_0$. \par
If the efficiency $Prob(s)$ is not too small, the frequency and their
variance in each bin should be affected by approximately the same
overall factor. Therefore the ratio of the numerator and denominator
of Eq. (1) should be only slightly dependent on $Prob(s)$. Furthermore,
the additional event--to--event fluctuations introduced by the
efficiency filtering is multinomial in character, and they
should be eliminated by the very definition of the factorial
moments [2]. This argumentation could be one possible
qualitative explanation for the stability of the signal,
observed in the numerical simulations. Similar results have
been obtained by assuming an efficiency $Prob(s)$ decreasing
with $s$. In particular we have studied the case with $Prob(1) = 1$
and $Prob(s)$ decreasing linearly with $s$ down to zero
for some value $s_0$ of $s$. For $s > s_0$, $Prob(s)$ was considered
to vanish identically. In this case apparent deviations from
linear rise was observed only for quite small values of $s_0$,
roughly for $s_0 < 10$.\par
We have also studied the effect of the mass/charge resolution of the
detector. In general the resolution does not produce any distortion
on the intermittency signal, even for unrealistically large values
of the resolution width. The mass/charge resolution uncertainty,
in fact, produces in each event a distortion of the mass/charge
distribution, but it cannot change the intrinsic fractal nature
of the distribution itself. This insensitivity to the resolution
was checked in explicit numerical simulations, where,
event by event, each fragment $s$--value was shifted randomly,
according to a normal distribution of a given width, before
calculating the factorial moments of Eq. (1).\par

The intermittency signal can be obscured or hidden not only by the
detector inefficiencies, but also by the dynamics itself of the
heavy--ion collisions. In particular, for non--central
collisions, it is possible that some part of the nuclear system is
only weakly excited. It forms  then the so--called
"spectator" part of the reaction. This subsystem can however still
emits a substantial fraction of the observed fragments,
thus obscuring the intermittency signal which is expected to be
originated by the "participating" part. This problem, however, can
be avoided if one selects central collisions and reactions
between nearly symmetrical systems.
\par
Another possible physical disturbance, due to the reaction dynamics,
is the presence of pre--equilibrium emitted particles, usually
nucleons. By pre--equilibrium particles we mean generically
particles that are emitted in the very preliminary stage of the
reactions. They are usually quite energetic and preferentially go in
the forward direction. At intermediate energies, according to
the theoretical simulations, one expects several pre--equilibrium
particles in all heavy--ion collisions. Their statistical
properties should be completely different with respect
to the ones of the particles emitted from the participating part.
This is particularly true if, in the participating
region, a phase transition occurs or an instability is present,
like in a genuine multifragmentation process, for which
intermittency should be present.
The simplest way of introducing the pre--equilibrium particles
in the previous schematic multifragmentation model is to add
to the percolating system another independent particle source.
The latter will be assumed to produce, on the average, a given number
$\overline {N}$ of particles ( $s$ = 1 ), with fluctuations
from one event to another which follow a gaussian distribution of
a given width $W$. After adding, for each event,
these particles to the fragment distribution produced by the
percolating system, we perform again the same intermittency
analysis and we check to which extent the signal is reduced.
The results show that the
disturbance depends mainly on the ratio $R = W/{\overline{N}}$, being
larger if the ratio is smaller. A typical example is reported in
Fig. 3, for $\overline{N} = 20$ and different $W$ values. For clarity
of the figure only $F_4$ is reported, but similar results are found
for the other moments. This value of $\overline{N}$ is about 25\%
of the average number of $s$=1 fragments produced by the percolating
system. When $R \approx 0.5$, namely the full width equal
the average value, the intermittency signal starts to degrade,
and it rapidly disappears for smaller values of R. This result
does not depend so much on $\overline{N}$ itself, except when
it is very small, of the order of few units, in which case the
effect of the additional source become trivially
negligible. The pre--equilibrium particles should be therefore
identified in experiments in which the intermittency
signal is looked for, and they should be systematically excluded
in the calculation of the factorial moments.
This is not always an easy task, since their energy spectra and
angular distributions partly overlap in general with the ones
pertinent to the system undergoing multifragmentation.\par
In summary we have presented a study of the disturbances on the
intermittency signal that can be present in experiments with
4$\pi$ detectors for heavy--ion collisions at intermediate energy.
The effects of both detector inefficiencies and reaction dynamics
have been considered. While most of the 4$\pi$ detectors,
that are already in operation, under construction,
or planned, are expected to have an efficiency good enough to be
sensitive to the intermittency signal, the main warning is about
the presence of pre--equilibrium particles. The latter could
cause, in fact, serious problems at the level of the phenomenological
analysis, since they could mask completely the intermittency signal
if they are not excluded in the calculation of the factorial
moments.\par
Valuable and stimulating discussions with M. Ploszajczak,
R. Pecshanki, and P. D\'esesquelles are kindly acknowledged.

\vfill\eject
\bigskip
\noindent
{\bf References}
\item{[1]} X.Campi, J.Phys. {\bf A19}(1986)917;\hfill\break
           X.Campi, Phys. Lett. {\bf B208}(1988)351.
\item{[2]} A.Bialas and R.Pecshanki, Nucl. Phys. {\bf B273}(1986)703.
\item{[3]} M.Plozsajczak and A.Tucholski, Phys. Rev. Lett.
{\bf 65}(1990)1539;
\hfill\break
           M.Plozsajczak and A.Tucholski, Nucl. Phys.
{\bf A523}(1991)651.
\item{[4]} C.J.Waddington and P.S.Freier, Phys. Rev.
 {\bf C31}(1988)888;
\hfill\break
           P.L.Juin {\it et al.}, Phys. Rev. Lett.
{\bf 68}(1992)1656.
\item{[5]} M.E.Brandan, A.J.Cole, P.D\'esesquelles, A.Giorni,
D.Heuer, A.LLeres, A.Mechaca--Rocha and K.Michaelian,
preprint ISN 92.57, (1992).
\item{[6]} H.R.Jaqaman and D.H.E.Gross, Nucl. Phys.
{\bf A524}(1991)321.
\hfill\break
           D.H.E.Gross, A.R.DeAngelis, H.R.Jaqaman, Pan Jicai and
and R.Heck, Phys. Rev. Lett. {\bf 68}(1992)146.
\item{[7]} D.Stauffer, Phys. Reports {\bf 54}(1979)1.
\item{[8]} W.Bauer, D.R.Dean, U.Mosel and U.Post, Nucl. Phys.
{\bf A452}(1986)699; H.Ngo, C.Ngo, F.Z.Ighezou, J.Debois,
S.Leray and Y.M.Zheng, {\bf A337}(1990)81; H.R.Jaqaman, G.Papp and
D.H.E.Gross, {\bf A514}(1990)327; L.Phair {\it et al.}, Phys. Lett.
{\bf B285}(1992)10; U.Lynen {\it et al.}, Nucl. Phys.
{\bf A545}(1992)329c.
\item{[9]} P.Finocchiaro, C.Agodi, R.Alba, G.Bellia, R.Coniglione,
 A.Del Zoppo, P. Maiolino, E.Migneco, G.Piattelli and P.Sapienza,
 preprint INFN--LNS, submitted to NIM. We thank Dr. P.Finocchiaro
for providing us the random number generator routine.
\item{[10]} W.A.Friedman, Phys. Rev. {\bf C46}(1992)R1595.
\item{[11]} W.A.Friedman, Phys. Rev. {\bf C42}(1990)667.
\vfill\eject
\bigskip
\bigskip
\noindent
{\bf Figure captions}
\bigskip
\par\noindent
\item{Fig. 1.} Scaled factorial moments for a $6^3$ percolating
lattice. The linear rise is observed only for case $b$),
for which the interval of random $q$ values includes the critical
value $q_c = 0.23$. This intermittency signal is the one used
in the present analysis.
\item{Fig. 2.} The intermittency signal of Fig. 1$b$), after being
filtered with the apparatus efficiency. For more details
see the text.
\item{Fig. 3.} The intermittency signal obtained by adding
pre--equilibrium particles (peps) to the source of Fig. 1$b$).
The quantity $W$ is the half--width of the peps number distribution.
For more details see the text.
\vfill\eject
\bye